\documentclass[twocolumn,notitlepage,prb,superscriptaddress,longbibliography]{revtex4-2}
\usepackage{mhchem}
\usepackage{amsthm}
\usepackage{amsmath}
\usepackage{amssymb}
\usepackage{mathdots}
\usepackage{graphicx}
\usepackage{mathrsfs}
\usepackage{longtable}
\usepackage{multirow}
\usepackage{xcolor}
\makeatletter

\@ifundefined{textcolor}{}
{%
 \definecolor{BLACK}{gray}{0}
 \definecolor{WHITE}{gray}{1}
 \definecolor{RED}{rgb}{1,0,0}
 \definecolor{GREEN}{rgb}{0,1,0}
 \definecolor{BLUE}{rgb}{0,0,1}
 \definecolor{CYAN}{cmyk}{1,0,0,0}
 \definecolor{MAGENTA}{cmyk}{0,1,0,0}
 \definecolor{YELLOW}{cmyk}{0,0,1,0}
}

\usepackage{braket}
\usepackage[backref=true,
bookmarksnumbered=true,
bookmarks=true,
bookmarksopen=true,
colorlinks=true,
citecolor=blue,
linkcolor=blue,
anchorcolor=green,
urlcolor=blue,unicode=false]{hyperref}

\let\baraccent=\= % rename builtin command \= to \baraccent
\renewcommand{\=}[1]{\stackrel{#1}{=}} % for putting numbers above =

\DeclareMathOperator{\Tr}{Tr}

\begin{document}

\title{Chiral Majorana hinge modes on a curved surface with magnetic impurities}
\author{Ilyoun Na}
    \email{ilyoun1214@berkeley.edu}
    \affiliation{Department of Physics, University of California, Berkeley, California 94720, USA}
    \affiliation{Materials Sciences Division, Lawrence Berkeley National Laboratory, Berkeley, California 94720, USA}
    \affiliation{Molecular Foundry, Lawrence Berkeley National Laboratory, Berkeley, CA 94720, USA}
\author{James G. McHugh}
    \affiliation{University of Manchester, School of Physics and Astronomy, Manchester M13 9PL, United Kingdom}
    \affiliation{National Graphene Institute, University of Manchester, Manchester M13 9PL, United Kingdom}
\author{Sin\'{e}ad M. Griffin}
    \affiliation{Materials Sciences Division, Lawrence Berkeley National Laboratory, Berkeley, California 94720, USA}
    \affiliation{Molecular Foundry, Lawrence Berkeley National Laboratory, Berkeley, CA 94720, USA}
\author{Luca Chirolli}
    \affiliation{Department of Physics and Astronomy, University of Florence, I-50019 Sesto Fiorentino, Italy}

\date{\today}

\begin{abstract}
Chiral Majorana modes, a key ingredient for topological quantum computation, arise on the surface of a three-dimensional time-reversal-invariant topological superconductor when a magnetic field induces a sign-changing surface mass. In practice, the orbital response of the superconductor tilts the surface spectrum and overwhelms the Zeeman term as the field grows. Here we show that magnetic impurities on a spherical surface can instead generate this mass term through their exchange coupling, so that only a weak field is required. The Ising surface Majorana modes mediate a ferromagnetic Ruderman-Kittel-Kasuya-Yosida (RKKY) interaction, derived here on the curved surface, that orders the radial impurity moments into an inversion-breaking hedgehog at zero field. An axial field instead stabilizes a domain wall across which the radial magnetization reverses sign, leaving a single chiral Majorana mode along the equator and realizing an inversion-protected second-order topological superconductor.
\end{abstract}

\maketitle

%====================================================================================================================================================================
%====================================================================================================================================================================

\textit{Introduction}.---Chiral one-dimensional Majorana modes are much sought after for their potential application in topological quantum computing~\cite{Nayak2008,Sarma2015,Chirolli2018,Lian2018,You2019,Bomantara2020,Pahomi2020}, and topological superconductors (TSCs)~\cite{Kitaev2001,Qi2011,Alicea2012,Flensberg2021,DasSarma2023} offer a promising platform for isolating these modes. A particularly direct route starts from a three-dimensional (3D) time-reversal-invariant (TRI) TSC, whose surface hosts a gapless Majorana cone. This parent phase is realized by superfluid $^3$He-B and is now being sought in UTe$_2$, where nonchiral odd-parity pairing~\cite{Gu2025}, an in-gap surface band~\cite{Wang2025,Gu2025}, and possibly a full gap~\cite{Suetsugu2024} have been reported. A time-reversal-breaking mass term that changes sign along a line on the surface binds a chiral Majorana mode to that line~\cite{Volovik2010,Fu2010}. In the language of higher-order topology~\cite{Song2017,Langbehn2017,Benalcazar2017,Schindler2018,Eslam2018,EslamSI2018,Zhu2018,Yang2019,Yang2020,Yang2024}, such a mode is the hinge mode of a second-order TSC.

The difficulty with this route is the field itself. In a bulk superconductor, Meissner screening produces an orbital supercurrent that couples to the surface Majorana modes and tilts their spectrum. As the field grows, this orbital effect dominates the Zeeman term~\cite{Chirolli2018_tilting,Chirolli2019}, so the sign-changing mass cannot be enlarged simply by increasing the field.

To address this, we propose decorating the surface with magnetic impurities. Their exchange coupling to the surface Majorana modes acts as a local Zeeman term, so aligning the moments with a weak external field enhances the mass term without enhancing the orbital supercurrent. The impurities, however, do not remain passive. The Ising character of the Majorana surface fermions~\cite{Chung2009} makes the Ruderman-Kittel-Kasuya-Yosida (RKKY) interaction they mediate
ferromagnetic on a planar surface~\cite{Choy2013}, as it is for the
Dirac surface states of a topological insulator~\cite{Abanin2011}, and
a dense ensemble of impurities orders on its own.

In this Letter we show that this self-ordering, rather than obstructing the proposal, provides its mechanism. We derive the RKKY interaction mediated by the Majorana modes on a sphere and find it ferromagnetic in the radial spin components. At zero field the impurities form a hedgehog resembling a magnetic monopole, which gaps the whole surface but breaks inversion symmetry. An axial field instead favors a domain-wall state in which the moments remain radial but reverse sign across the equator. The transition is governed by the competition between the extensive Zeeman energy gain, $\propto NB$ for $N$ impurities in a field with Zeeman energy $B$ per moment, and the sub-extensive cost of the wall, $\propto J\sqrt{N}$ with $J$ the nearest-neighbor RKKY coupling, so that the domain-wall state becomes the ground state at a critical field $B_c\propto J\,d/R$, with $d$ the impurity spacing and $R$ the radius. For fields of order $J$, the moments cant continuously toward the field, with no further transition.

The field is thus only required to select the domain-wall configuration, whereas the mass term that gaps the surface is supplied by the exchange coupling to the ordered impurities. This is what allows a single gapless chiral Majorana mode to be stabilized along the equator at fields well below the scale at which orbital effects dominate. Because inversion symmetry is preserved, this realizes the nontrivial element of the $\mathbb{Z}_2$ classification of inversion-symmetric second-order TSCs in class D~\cite{Eslam2018,Luka2019}.

\begin{figure}[t]
    \centering
    \includegraphics[width=\columnwidth]{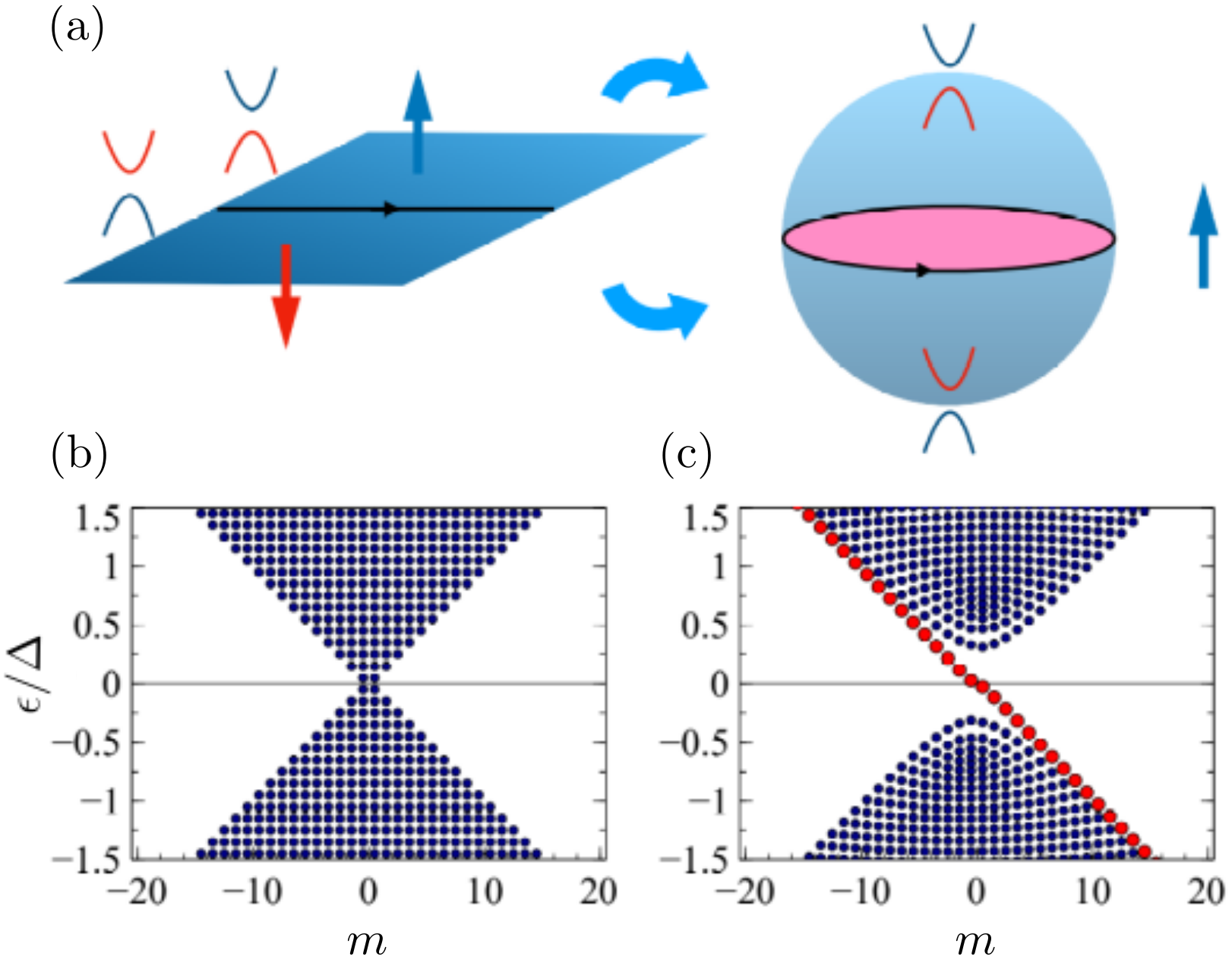}
    \caption{(a) A mass term that changes sign across a line on a planar
    surface (left) binds a chiral Majorana mode to that line. On a sphere
    (right), a uniform Zeeman field along the polar axis produces such a
    mass, $V_\perp=V\cos\theta$, which vanishes on the inversion-symmetric
    equator. (b) Surface spectrum of Eq.~\eqref{eq:HdiracS} as a function of
    the polar angular momentum $m$, without and (c) with the Zeeman field.
    The branch crossing the gap in (c) is the chiral equatorial mode.}
    \label{fig:espectrum}
\end{figure}

\textit{Second-order TSC with Zeeman field}.---On a planar surface, the Majorana cone of a 3D TRI odd-parity class DIII TSC~\cite{Ryu2010,Fu2010,Chiu2016} is described by $H=-i\hbar v_\Delta(\partial_x\sigma_y-\partial_y\sigma_x)$, where the Pauli matrices $\sigma_i$ act on the surface Kramers doublet in the Majorana representation~\cite{SuppMat} and $v_\Delta=\Delta/\hbar k_F$. Projected onto this doublet, the in-plane spin components vanish while $s_z$ maps onto $\sigma_z$, the Ising character of the surface Majorana modes~\cite{Chung2009}, so a Zeeman field with normal component $V_z$ enters as the mass term $V_z\sigma_z$. For $V(y)={\rm sgn}(y)V_z$, a chiral mode propagates along $x$, localized at the node of the mass. On a sphere, a uniform Zeeman field along the polar axis provides such a mass, whose normal component vanishes on the equator where the chiral mode resides, as shown in Fig.~\ref{fig:espectrum}(a).

\textit{Surface Majorana modes on a sphere}.---We now place the surface Hamiltonian on a sphere of radius $R$, following Refs.~\cite{Gonzalez1992,AbrikosovJr,Parente2011,Chirolli2018_tilting}. In terms of the flat-space gamma matrices $(\gamma^0,\gamma^1,\gamma^2)=(-i\sigma_z,\sigma_x,\sigma_y)$, $\{\gamma^i,\gamma^j\}=2\eta^{ij}$, $\eta_{ij}={\rm diag}(-1,1,1)$, the planar Hamiltonian reads $H=\hbar v_\Delta\,\gamma^0(-i\gamma^a\partial_a)$, where $a=1,2$, so that $i\hbar\partial_t\psi=H\psi$ is the massless Dirac equation. On a curved surface with metric $g_{\mu\nu}$, the gamma matrices become position dependent, $\gamma^\mu(x)=\gamma^ie_i^{~\mu}$~\footnote{The tetrad $e^i_{~\mu}\equiv\partial x^i/\partial x^\mu$ relates the two frames, $g_{\mu\nu}=\eta_{ij}e^i_{~\mu}e^j_{~\nu}$, and $e_i^{~\mu}$ is its inverse. The spin connection is $\Gamma_\mu=\frac{i}{2}\Gamma^{i~~j}_{~\mu}\Sigma_{ij}$ with $\Sigma_{ij}=\frac{i}{2}[\gamma_i,\gamma_j]$ and $\Gamma^{i~~j}_{~\mu}=e^i_{~\nu}\nabla_\mu e^{j\nu}$.} with $\{\gamma^\mu(x),\gamma^\nu(x)\}=2g^{\mu\nu}$, and $\partial_a$ is replaced by the covariant derivative $D_a=\partial_a+\Gamma_a$, with $\Gamma_a$ the spin connection~\cite{Wald1984,Nakahara1990}. For the sphere, $g_{\mu\nu}={\rm diag}(-1,R^2,R^2\sin^2\theta)$, so $\gamma^0=-i\sigma_z$, $\gamma^1(x)=\sigma_x/R$, $\gamma^2(x)=\sigma_y/(R\sin\theta)$, and $\Gamma_a=-\frac{i}{2}\sigma_z\cos\theta\,\delta_{a,2}$, giving
\begin{equation} \label{eq:HdiracS}
    h_0^{\rm cur}=\frac{\Delta}{k_F R}~\gamma^0 \left[-i\sigma_x \left(\partial_\theta+\frac{\cot\theta}{2}\right) - i\frac{\sigma_y}{\sin\theta}~\partial_\phi~\right].
\end{equation}
The natural basis for Eq.~\eqref{eq:HdiracS} is the set of eigenstates $|\Upsilon^\sigma_{lm}\rangle$ of the spatial Dirac operator $-i\gamma^aD_a$, with eigenvalues $\sigma(l+1/2)/R$, $\sigma=\pm1$~\cite{AbrikosovJr,SuppMat}. They are given by Jacobi polynomials and labeled by the half-integer angular momentum $l$ and its polar projection $m$. Because $\gamma^0=-i\sigma_z$ flips $\sigma$, $h_0^{\rm cur}$ is off-diagonal in this basis, and its eigenstates are $|\Psi^\pm_{lm}\rangle=(|\Upsilon^+_{lm}\rangle\mp i|\Upsilon^-_{lm}\rangle)/\sqrt2$ with energies $\pm(l+1/2)\Delta/(k_FR)$, each $(2l+1)$-fold degenerate~\cite{AbrikosovJr,Zhang2010,Imura2012,Takane2013}. The curvature thus gaps the Majorana cone, with a gap $2\Delta/(k_FR)$ that closes as $R\to\infty$, as shown in Fig.~\ref{fig:espectrum}(b).

On the sphere the surface normal is the radial direction, so a uniform field $\mathbf V$ along the polar axis enters Eq.~\eqref{eq:HdiracS} as $h_Z=V_\perp\sigma_z$ with $V_\perp=V\cos\theta$. Since the field is an axial vector it is even under inversion, whereas the outward normal is odd, so $V_\perp(\mathbf r)=-V_\perp(-\mathbf r)$. The mass therefore changes sign across the inversion-symmetric equator, $\theta=\pi/2$, where the field is tangential to the surface, and inversion symmetry is preserved.

In the $\Upsilon$ basis, the Zeeman term has off-diagonal elements connecting $l'=l\pm1$,
\begin{align}
    \langle\Upsilon^+_{lm}|\sigma_z\cos\theta|\Upsilon^-_{l'm'}\rangle=&~i\delta_{m,m'}\frac{\sqrt{(l+1)^2-m^2}}{2(l+1)}
    \delta_{l',l+1} \label{eq:matrix}  \\
    &-i\delta_{m,m'}\frac{\sqrt{l^2-m^2}}{2l}\delta_{l',l-1}, \nonumber
\end{align}
which open an additional gap, and diagonal elements~\cite{Chirolli2018_tilting},
\begin{equation} \label{eq:diagonal}
    \langle\Upsilon^\pm_{lm}|\sigma_z\cos\theta|\Upsilon^\pm_{l'm'}\rangle=\delta_{m,m'}\delta_{l,l'}\frac{m}{2l(l+1)},
\end{equation}
which are the same in the $\Psi$ basis. Diagonalizing $h_0^{\rm cur}+h_Z$ in the $\Upsilon$ basis gives Fig.~\ref{fig:espectrum}(c). The diagonal elements shift all states with positive $m$ upward and all states with negative $m$ downward, most strongly at small $l$. For $l=|m|$ the shift undoes the gapping, leaving a single chiral mode crossing the gap.

\textit{RKKY interaction among magnetic impurities on a sphere}.---To obtain the mass term at reduced field, and hence reduced orbital tilting~\cite{Chirolli2018_tilting,Chirolli2019}, we now place magnetic impurities $\mathbf S_i$ at positions $\mathbf R_i$ on the surface, treated as classical spins coupled to the Majorana modes by a local exchange interaction. Because only the spin component normal to the surface couples, the exchange term on the sphere involves only the radial component $S_i^r$ of each moment,
\begin{equation} \label{eq:h_RKKY}
    h_{\rm ex}^{\rm cur}= i\gamma^0\lambda \sum_{i} S_i^{r}\,\delta(\mathbf x-\mathbf R_i)\frac{1}{\sqrt{|g|}},
\end{equation}
with $i\gamma^0=\sigma_z$ and $\lambda$ the exchange strength. The factor $|g|^{-1/2}$, with $g$ the determinant of $g_{\mu\nu}$, is removed by the rescaling $\Psi\to|g|^{1/4}\Psi$, after which the spinors obey flat-space anticommutation relations~\cite{SuppMat}.

\begin{figure}[t]
    \centering \includegraphics[width=\columnwidth]{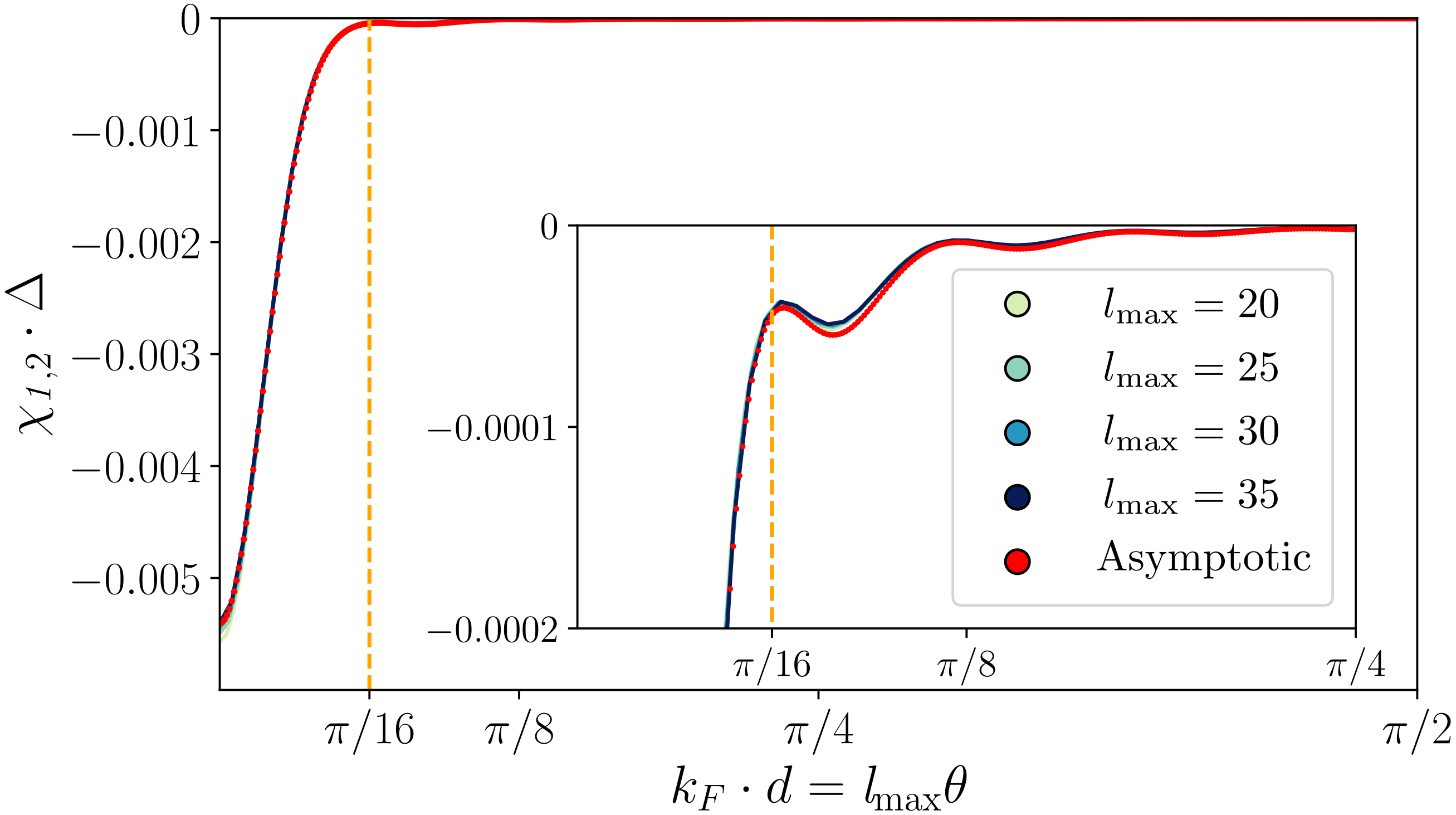}
    \caption{The dimensionless static susceptibility $\chi_{1,2}\Delta$ of the surface modes, with lengths in units of $k_F^{-1}$, versus the dimensionless arc distance $k_Fd=k_FR\theta=l_{\rm max}\theta$ between two impurities, one at the north pole and one at polar angle $\theta$, with $l_{\rm max}=k_FR$ the cutoff in Eq.~\eqref{eq:susceptiblity}. $\chi_{1,2}<0$ at all separations, so the RKKY coupling $\lambda^2\chi_{1,2}$ is ferromagnetic in the radial components. The inset shows the oscillatory tail for several $l_{\rm max}$, compared with the asymptotic form of Eq.~\eqref{eq:RKKY asymptotic}.}
    \label{fig:RKKY_oscillation}
\end{figure}

The impurities interact through the surface Majorana modes. Integrating out the Majorana modes and expanding the resulting effective action to second order in $\lambda$ yields the RKKY interaction
\begin{equation} \label{eq:H_RKKY}
    H^{\rm RKKY}=\lambda^2\sum_{i,j}\chi_{i,j}S^{r}_i S^{r}_j,
\end{equation}
where $\chi_{i,j}(\tau)=\langle T_\tau s^{r}(\mathbf R_i,\tau)\,s^{r}(\mathbf R_j,0)\rangle$ is the radial spin susceptibility of the surface modes, with $T_\tau$ the imaginary-time ordering operator, and $\chi_{i,j}\equiv\chi_{i,j}(i\nu_n=0)$ is its static limit, with $\chi_{i,j}(i\nu_n)=\int_0^\beta d\tau\,e^{i\nu_n\tau}\chi_{i,j}(\tau)$, bosonic Matsubara frequencies $\nu_n=2n\pi/\beta$, and $\beta=1/T$ ($k_B\equiv1$). At one loop, the fermion loop gives $\chi_{i,j}(i\nu_n)=-\frac{1}{\beta}\sum_{iw_n}\Tr[s^{r}_i\,G_0(\mathbf R_i,\mathbf R_j,i\nu_n+iw_n)\,s^{r}_j\,G_0(\mathbf R_j,\mathbf R_i,iw_n)]$, with fermionic Matsubara frequencies $w_n=(2n+1)\pi/\beta$~\cite{Kapusta1989}, $G_0=(iw-h_0^{\rm cur})^{-1}$ the Green's function of Eq.~\eqref{eq:HdiracS}, and the trace running over its eigenstates $|\Psi^\pm_{lm}\rangle$~\cite{SuppMat}. In the low-temperature limit $T\ll\Delta/(k_FR)$, the static susceptibility reduces to
\begin{equation} \label{eq:susceptiblity}
    \chi_{i,j}=-2\sum_{lm,l'm'}\frac{Z_{lm;l'm'}(\mathbf R_j)Z^*_{lm;l'm'}(\mathbf R_i)}{\mathcal E_l+\mathcal E_{l'}},
\end{equation}
where $Z_{lm;l'm'}(\mathbf R)=\Psi^{+\dagger}_{lm}(\mathbf R)\sigma_z\Psi^-_{l'm'}(\mathbf R)$ connects positive and negative energy states, $\mathcal E_l=(l+1/2)\Delta/(k_FR)$, and the sum runs over $|m|\le l\le l_{\rm max}+1/2$, where the integer $l_{\rm max}\simeq k_FR$ marks the angular momentum at which $\mathcal E_l$ reaches the bulk gap $\Delta$. Using azimuthal symmetry, we fix one impurity at the north pole and the other at polar angle $\theta$, and plot the dimensionless $\chi_{1,2}\Delta$ against the dimensionless arc distance $k_Fd=k_FR\theta=l_{\rm max}\theta$ in Fig.~\ref{fig:RKKY_oscillation}. The coupling $\lambda^2\chi_{1,2}$ is negative at all separations, so the RKKY interaction is ferromagnetic in the radial components. Below we denote the nearest-neighbor coupling by $J\equiv-\lambda^2\chi_{1,2}(d)>0$, evaluated at the impurity spacing $d$.

To obtain its asymptotic form, we place both impurities on the equator at azimuthal separation $\phi$, equivalent to the geometry of Fig.~\ref{fig:RKKY_oscillation} by rotational symmetry, and take $l_{\rm max}$ large and $\phi\ll\pi$~\cite{SuppMat}, which gives
\begin{equation} \label{eq:RKKY asymptotic}
    \chi(\phi)=-\frac{k_F}{4\pi^2 R^3\Delta}\sum_{l,l'=1/2}^{l_{\rm max}+1/2} \frac{(l+1/2)(l'+1/2)}{l+l'+1}{\cal F}_{l,l'}(\phi),
\end{equation}
with ${\cal F}_{l,l'}(\phi)=J_0(l\phi)J_{0}(l'\phi)+J_1(l\phi)J_{1}(l'\phi)$.
It agrees with the exact result for $l_{\rm max}>20$ (Fig.~\ref{fig:RKKY_oscillation}, inset) and reduces to the flat-surface result~\cite{Choy2013} as $R\to\infty$.

Since the RKKY interaction is ferromagnetic in the radial components, the impurities are expected to order radially. Such order gaps the surface Majorana modes, and the gap $\Delta_{\rm s}$ in turn cuts off the range of the RKKY interaction at a length $\sim\hbar v_\Delta/\Delta_{\rm s}$. A self-consistent treatment of the gap and the magnetization is beyond the scope of this work. We therefore restrict the phase-diagram analysis below to a nearest-neighbor interaction, with $J$ set by Eq.~\eqref{eq:susceptiblity}, and expect longer-range couplings to modify prefactors but not the extensive-versus-sub-extensive competition that determines the transition. 

\textit{Magnetic phase diagram}.---We parametrize each moment by the angle $\Theta_i$ between $\mathbf S_i$ and the outward normal $\hat r_i$ at $\mathbf R_i=(\theta_i,\phi_i)$. Since the RKKY term involves only $S_i^r=S\cos\Theta_i$ and the Zeeman term only $\hat z\cdot\mathbf S_i$, the energy is minimized with each moment in the plane of $\hat r_i$ and the meridian tangent $\hat\theta_i$, $\mathbf S_i=S(\cos\Theta_i\,\hat r_i+\sin\Theta_i\,\hat\theta_i)$, for which $\hat z\cdot\mathbf S_i=S\cos(\Theta_i+\theta_i)$ because $\hat z=\cos\theta_i\,\hat r_i-\sin\theta_i\,\hat\theta_i$. With the nearest-neighbor interaction of Eq.~\eqref{eq:H_RKKY} the energy then reads
\begin{equation} \label{Eq:Hij}
    H = -J\sum_{\langle ij\rangle}\cos\Theta_i\cos\Theta_j-B\sum_j\cos(\Theta_j+\theta_j),
\end{equation}
where $J>0$ was defined above, $B$ is the Zeeman energy of a moment aligned with the field, and the spin magnitude $S$ is absorbed into $J$ and $B$. The ferromagnetic RKKY term favors radial alignment, $\Theta_i=0$ or $\pi$, while the Zeeman term favors the axial configuration $\Theta_i=-\theta_i$. The ground state follows from the competition between the two.

\begin{figure}[t]
    \centering
    \includegraphics[width=\columnwidth]{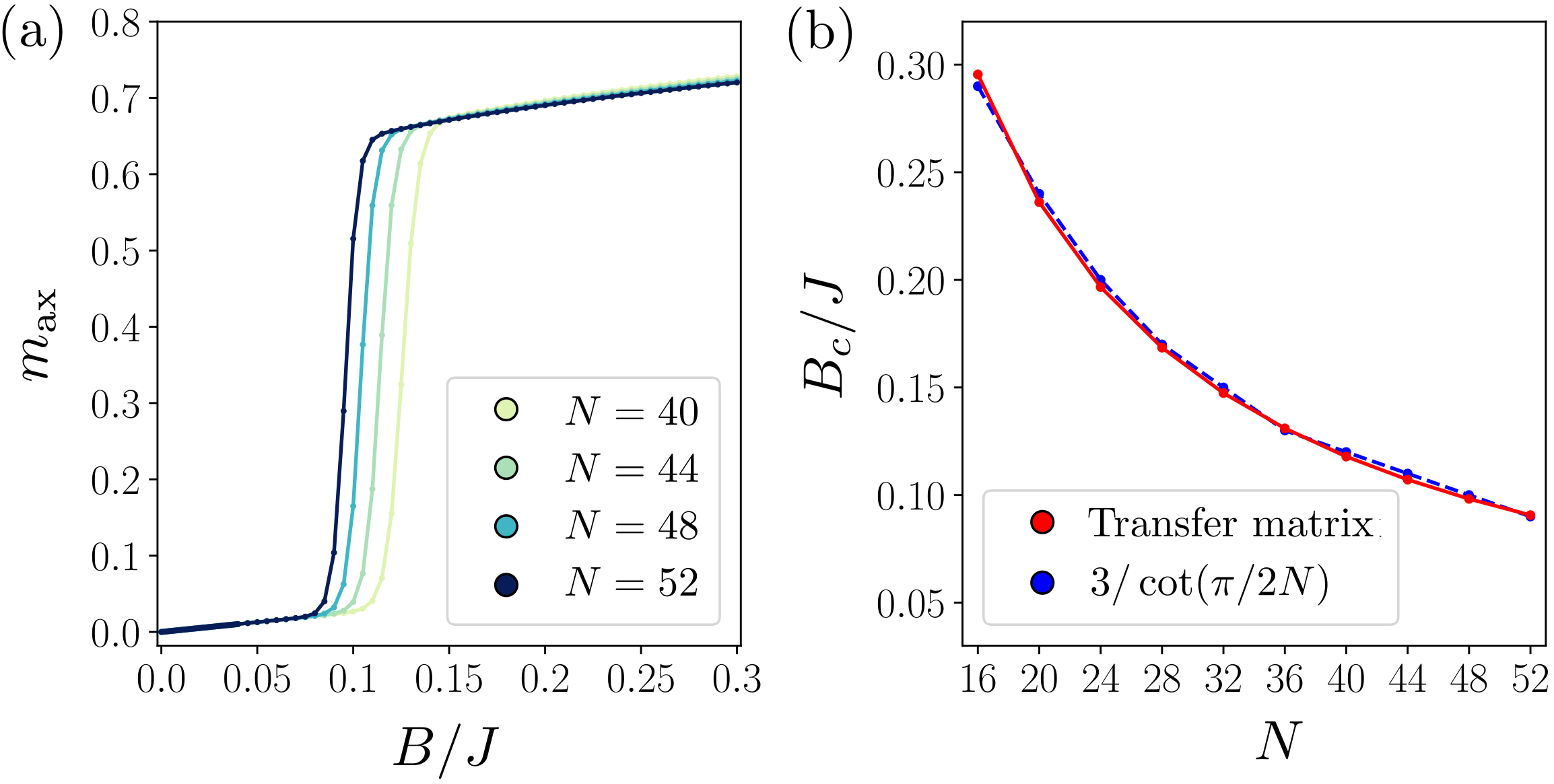}
    \caption{(a) Axial magnetization $m_{\rm ax}$ of $N$ impurities on a great circle, from the transfer-matrix evaluation of Eq.~\eqref{Eq:Hij} with $M=100$ and $T/J=0.1$. (b) Critical field $B_c$ of the transition from the hedgehog to the domain-wall configuration, extracted from (a), versus $N$. The dashed line is $B_c/J=3/\cot(\pi/2N)$, which follows the $1/N$ scaling of the zero-temperature energy balance, $B_c/J=4/\cot(\pi/2N)$, with a smaller prefactor.
    \label{fig:magnetization}}
\end{figure}

\begin{figure}[t]
    \centering
    \includegraphics[width=\columnwidth]{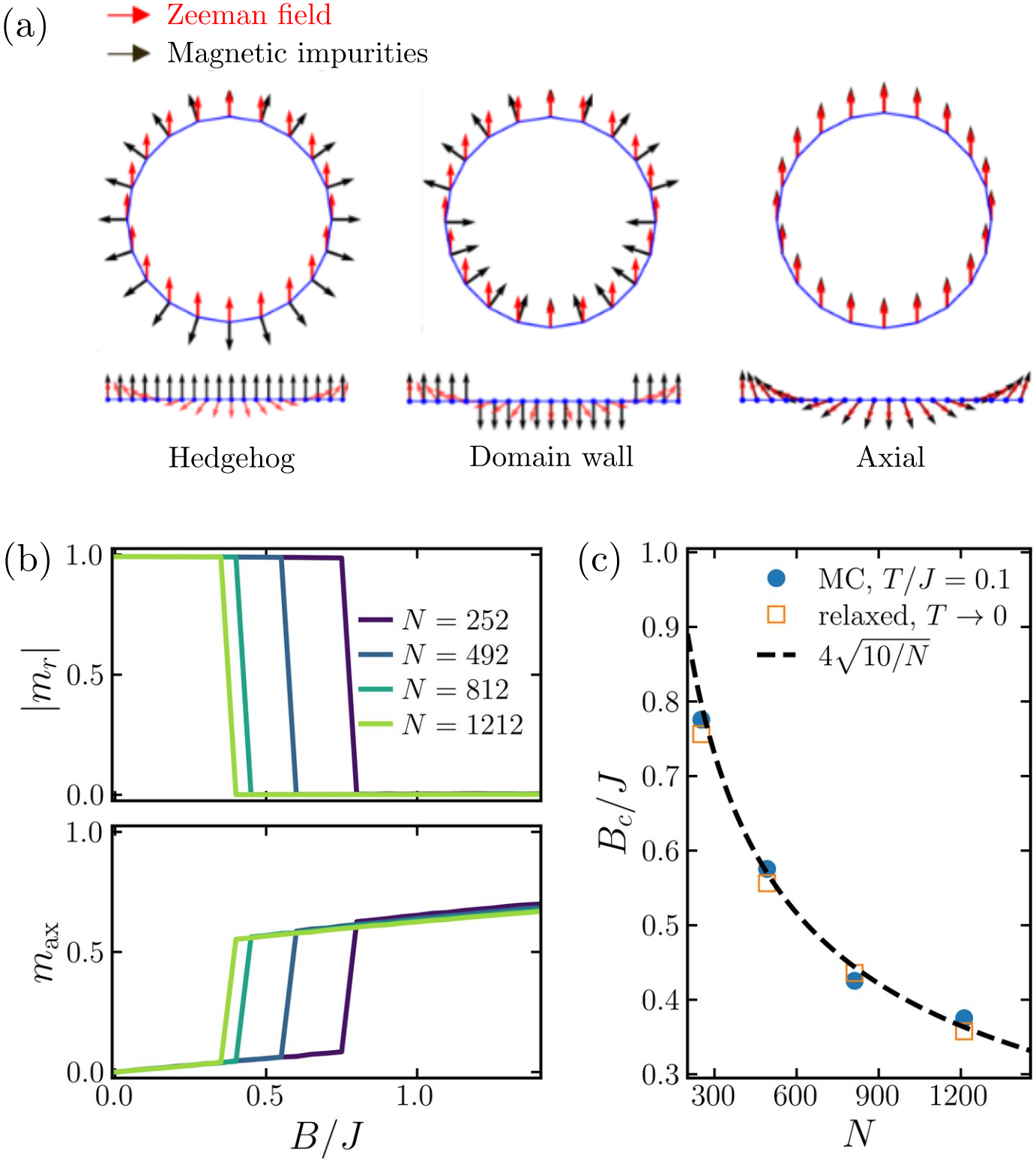}
    \caption{(a) Impurity moments (black) and field (red) along a geodesic in the hedgehog, domain-wall, and axial configurations. (b) Monte Carlo results for Eq.~\eqref{Eq:Hij} on the icosahedral sphere at $T/J=0.1$ for $N=252$ to $1212$, showing the radial magnetization $|m_r|$ (top) and the axial magnetization $m_{\rm ax}$ (bottom) versus $B/J$. (c) Critical field of the hedgehog to domain-wall transition from (b) and from zero-temperature relaxation of the two configurations, compared with $4J\sqrt{10/N}$.
    \label{fig:Configuration}}
\end{figure}

Before treating the sphere, it is instructive to consider the one-dimensional problem of $N$ equally spaced impurities, $\theta_i=2\pi i/N$, on a great circle, a geodesic of the sphere, which already exhibits the competition between the three configurations of Fig.~\ref{fig:Configuration}(a). In the hedgehog, $\Theta_i=0$ for all $i$, every bond is satisfied, $E_{\rm hed}=-NJ$, and the Zeeman contributions of antipodal impurities cancel. In the axial configuration, $E_{\rm ax}=-NJ\cos(2\pi/N)/2-NB$. In the domain-wall configuration, $\Theta_i=0$ on the half circle with $\cos\theta_i>0$ and $\Theta_i=\pi$ on the other, only the two bonds crossing the wall are frustrated, and $E_{\rm dw}=-(N-4)J-B\cot(\pi/2N)$. Equating $E_{\rm dw}$ with $E_{\rm hed}$, the domain wall becomes the ground state when the extensive Zeeman gain outweighs the $N$-independent cost of the two frustrated bonds, $B>B_c=4J/\cot(\pi/2N)$, which for large $N$ gives $B_c\simeq2\pi J/N$. Equating $E_{\rm dw}$ with $E_{\rm ax}$, using $\cot(\pi/2N)\simeq2N/\pi$ and $\cos(2\pi/N)\simeq1$ for large $N$, the domain wall remains the lowest of the three configurations up to
$B\simeq J/[2(1-2/\pi)]\simeq1.4J$. The critical field thus decreases as $1/N$ at fixed $J$.

To test this estimate at finite temperature, we evaluate the partition function $Z=\int\prod^N_{i=1}\frac{d\Theta_i}{2\pi}\,e^{-H[\{\Theta_i\}]/T}\equiv e^{-F/T}$, keeping the moments in the meridian plane, by discretizing $\Theta_i=2\pi n_i/M$ and using the transfer-matrix method~\cite{SuppMat}, and compute the axial magnetization $m_{\rm ax}=-\frac1N\partial F/\partial B$. Figure~\ref{fig:magnetization}(a) shows $m_{\rm ax}$ versus $B$ at $T/J=0.1$. The magnetization jumps from zero to a finite value at a critical field that decreases with $N$ as $\simeq3/\cot(\pi/2N)$, as shown in Fig.~\ref{fig:magnetization}(b), the same $1/N$ scaling as the zero-temperature estimate, with the prefactor reduced by thermal fluctuations.

Turning to the sphere, we place $N$ impurities on the vertices of an icosahedral triangulation~\cite{SuppMat}, with $N_e=3(N-2)$ edges and $N_t=2(N-2)$ triangles. Equating the triangle area $4\pi R^2/N_t$ to $\sqrt3d^2/4$ gives the nearest-neighbor arc angle $\theta_d=d/R=\sqrt{16\pi/(\sqrt3N_t)}\simeq3.8/\sqrt N$, at which $J$ is evaluated. The aim is a qualitative phase diagram, not an exact account of each impurity position, so sums over the uniformly distributed sites are replaced by sphere averages, $\sum_jf(\theta_j)\simeq N\langle f\rangle$ with $\langle f\rangle=\frac12\int_0^\pi f(\theta)\sin\theta\,d\theta$, where only $\theta_j$ enters because the configurations and the field are axially symmetric. The three configurations are defined as on the great circle. In the hedgehog, $\Theta_i=0$, every bond is satisfied and the Zeeman term averages to zero. In the axial configuration, $\Theta_i=-\theta_i$, neighboring moments have $\cos\Theta_i\cos\Theta_j\simeq\cos^2\theta_i$ with $\langle\cos^2\theta\rangle=1/3$. In the domain-wall configuration, $\Theta_i=0$ for $\cos\theta_i>0$ and $\Theta_i=\pi$ for $\cos\theta_i<0$, the $\sqrt{10N}$ bonds crossing the equator~\cite{SuppMat} are frustrated, and the Zeeman energy is $-B\sum_j|\cos\theta_j|\simeq-NB\langle|\cos\theta|\rangle=-NB/2$. Hence
\begin{equation} \label{eq:E_sphere}
\begin{aligned}
E_{\rm hed}&=-N_eJ=-3(N-2)J,\\
E_{\rm ax}&\simeq-N_eJ/3-NB\simeq-NJ-NB,\\
E_{\rm dw}&\simeq-\big(3N-2\sqrt{10N}\big)J-NB/2.
\end{aligned}
\end{equation}
The domain wall is the ground state for $B>B_c=4J\sqrt{10/N}\simeq3.3\,J\,d/R$, and up to $B\simeq4J(1-\sqrt{10/N})$, beyond which the axial configuration has the lowest energy of the three. Compared with the great circle, the domain wall now costs $\propto J\sqrt N$ rather than a constant, so $B_c$ scales as $1/\sqrt N$ at fixed $J$.

We test this picture by classical Monte Carlo simulations of Eq.~\eqref{Eq:Hij} on the icosahedral triangulation with for $N=252$, $492$, $812$, and $1212$, the sizes for which no vertex lies on the equator. Over this range $B_c$ is predicted to decrease by a factor $\sqrt5$, which tests the $1/\sqrt N$ scaling. Besides single-moment Metropolis updates, the simulations include global moves that reverse all moments on one hemisphere, so that the hedgehog and the domain wall are sampled on an equal footing rather than selected by the initial state~\cite{SuppMat}. Figure~\ref{fig:Configuration}(b) shows the axial magnetization $m_{\rm ax}$ and the magnitude of the radial magnetization, $|m_r|$ with $m_r=\frac{1}{NS}\sum_i\langle S_i^r\rangle$, at $T/J=0.1$. At $B_c$, $|m_r|$ drops from one to zero and $m_{\rm ax}$ jumps from zero to $\simeq1/2$, the value $\langle|\cos\theta|\rangle$ of the domain-wall state. Figure~\ref{fig:Configuration}(c) compares this $B_c$ with the field at which the energies of the hedgehog and the domain wall cross when each configuration is relaxed at $T\to0$, so that the moments cant toward the field but the configuration is kept. Both follow $4J\sqrt{10/N}$ within $5\%$ for all sizes. Beyond the range of Fig.~\ref{fig:Configuration}(b), the simulations show no further transition up to $B=4.5J$~\cite{SuppMat}. The moments cant continuously toward the field and the sharp wall evolves into the axial texture $S^r\simeq S\cos\theta$, which also reverses sign at the equator, so the crossing of $E_{\rm dw}$ and $E_{\rm ax}$ at $B\simeq4J(1-\sqrt{10/N})$ is a crossover rather than a phase boundary.

These scalings hold at fixed $J$. At fixed exchange $\lambda$ and radius $R$, adding impurities reduces $d$, and $J=-\lambda^2\chi_{1,2}(d)$ responds according to Fig.~\ref{fig:RKKY_oscillation}. There
$|\chi_{1,2}|$ is appreciable only for $k_Fd\lesssim\pi/16$, where it increases steeply as $d$ decreases, and is two orders of magnitude smaller beyond. In this regime $J$ grows with $N$ faster than the geometric factor $d/R$ shrinks, so at fixed $\lambda$ and $R$ the critical field $B_c=4J(d)\sqrt{10/N}$ increases with $N$ rather than decreasing.

Independently of how $N$ is varied, $B_c\propto J(d)\,d/R$. Two consequences follow. First, at fixed impurity density $n=N/4\pi R^2$, the spacing $d\propto n^{-1/2}$ and hence $J(d)$ are fixed, so $B_c$ decreases as $1/R$. Second, since $d\ll R$, $B_c$ is much smaller than the RKKY scale $J$. The reason is that the domain wall frustrates only the bonds crossing the equator, a fraction of order $d/R$ of all bonds, while the field acts on every impurity. Finally, the radial magnetization of the domain-wall state, $S\,{\rm sgn}(\cos\theta)$, is the inversion-symmetric sign-changing mass of Fig.~\ref{fig:espectrum}, so a chiral Majorana mode appears along the equator, as in Fig.~\ref{fig:espectrum}(c), with a gap $\lambda S$ set by the exchange coupling rather than by the field.

\textit{Conclusion and outlook}.---We have shown that magnetic impurities on the surface of a 3D TRI odd-parity TSC can generate the sign-changing mass needed for a chiral Majorana hinge mode, with the external field playing only a selecting role. The Ising character of the surface Majorana modes makes their RKKY interaction ferromagnetic in the radial components. On a sphere this ordering yields a hedgehog at zero field and an inversion-symmetric domain-wall state above a critical field $B_c\propto J\,d/R$, set by the competition between the extensive Zeeman gain and the sub-extensive cost of the wall, as confirmed by Monte Carlo simulations. In the domain-wall state the surface gap is set by the exchange coupling rather than by the field, which keeps the required field well below the scale at which orbital effects dominate, and the equator hosts a single chiral Majorana mode, the nontrivial element of the $\mathbb{Z}_2$ classification of inversion-symmetric second-order TSCs in class D. The mode persists at larger fields, where the moments cant continuously toward the field without further transition. Two extensions follow naturally. A self-consistent treatment of the surface gap and the impurity order would determine the effective range of the RKKY interaction and refine the phase boundaries. The equatorial mode also offers a setting for interferometry~\cite{Li2021,Chaou2023}. UTe$_2$ could provide a material platform if its bulk superconducting state realizes the fully gapped, time-reversal-invariant odd-parity phase proposed in Ref.~\cite{Suetsugu2024}. The sphere stands in for any closed surface on which the moments can order along the local normal, so
a small single crystal or microparticle decorated with magnetic adatoms, in a weak field along one of its axes, would be the natural setting.

\textit{Acknowledgments.}---This work (I.N., S.M.G.) was supported by the Theory of Materials Program (KC2301) funded by the US Department of Energy, Office of Science, Basic Energy Sciences, Materials Sciences and Engineering Division under Contract No. DE-AC02-05CH11231. Work at the Molecular Foundry was supported by the Office of Science, Office of Basic Energy Sciences, of the US Department of Energy under the same contract No. Computational resources were provided by the National Energy Research Scientific Computing Center (NERSC). This project (L.C.) has received funding from the European Union’s Horizon 2020 research and innovation program under the Marie Sk{\l}odowska-Curie Grant Agreement No. 841894 (TOPOCIRCUS). J.G.M. is supported by the University of Manchester Dame Kathleen Ollerenshaw Fellowship.
%This work (I.N., S.M.G.) was supported by the US Department of Energy, Office of Science, National Quantum Information Science Research Centers, Quantum Systems Accelerator (QSA). Computational resources were provided by the National Energy Research Scientific Computing Center and the Molecular Foundry, DOE Office of Science User Facilities supported under Contract No. DEAC02-05-CH11231. 

\bibliography{reference}

%====================== Supplemental Material ======================
\clearpage
\onecolumngrid
\begin{center}
{\large\bfseries Supplemental Material for: \\[3pt] Chiral Majorana hinge modes on a curved surface with magnetic impurities} 
\end{center}
\vspace{9pt}
\twocolumngrid

\setcounter{section}{0}
\setcounter{equation}{0}
\setcounter{figure}{0}
\renewcommand{\theequation}{S\arabic{equation}}
\renewcommand{\thefigure}{S\arabic{figure}}

\section{DIII topological superconductor with magnetic impurities}

We consider the Balian-Werthamer (BW) state~\cite{BalianWerthamer}, the pairing state of superfluid $^3$He-B and the simplest representative of class DIII topological superconductors (TSCs). For a spin-degenerate three-dimensional parabolic band it is a triplet state with pairing $(\mathbf p\cdot\mathbf s)\,is_y$. In the Nambu basis $\psi_{\mathbf k}=(c_{\mathbf k\uparrow},c_{\mathbf k\downarrow}, c^\dagger_{-\mathbf k\downarrow},-c^\dagger_{-\mathbf k\uparrow})^T$, with $c_{\mathbf ks}$ annihilating an electron of momentum $\mathbf k$ and spin $s$, the Bogoliubov-de Gennes Hamiltonian reads
\begin{equation}
    H_0=\tau_z\left(\frac{p^2}{2m^*}-\mu\right)+\frac{\Delta}{\hbar k_F}\,\tau_x\,\mathbf p\cdot\mathbf s,
\label{eq:BW}
\end{equation}
where $m^*$ is the effective mass, $\mu$ the chemical potential, $k_F=\sqrt{2m^*\mu}/\hbar $ the Fermi momentum, $\mathbf s$ the vector of spin Pauli matrices, $\tau_i$ the Pauli matrices in Nambu space, and $\Delta$ the mean-field order parameter, which opens a full gap $\Delta$ on the Fermi surface. Magnetic impurities $\mathbf S_i$ at positions $\mathbf R_i$ on the surface couple to the electrons through a local exchange interaction,
\begin{equation}
    H_{\rm ex}=\lambda\sum_i\mathbf S_i\cdot\mathbf s\,\delta(\mathbf r-\mathbf R_i),
\label{eq:Hex}
\end{equation}
with exchange strength $\lambda$. The impurity spins are treated as classical vectors of fixed magnitude $S$.  

\subsection{Surface Majorana modes on a planar surface}

The nontrivial topology of Eq.~\eqref{eq:BW} produces a Majorana cone on the surface. For a superconductor occupying the half space $z>0$, a zero-energy Majorana Kramers pair is found at $k_x=k_y=0$, with wavefunctions
\begin{equation}
    \phi_s(z)=|s\rangle\big(|+\rangle_\tau+is|-\rangle_\tau\big)\,e^{-z/\xi}\sin(k_Fz),
\label{eq:phis}
\end{equation}
where $\xi=\hbar v_F/\Delta$ is the coherence length, $v_F=\hbar k_F/m^*$, $\tau_z|\pm\rangle_\tau=\pm|\pm\rangle_\tau$, and $s_z|s\rangle=s|s\rangle$ with $s=\pm1$. Projecting Eq.~\eqref{eq:BW} onto the doublet $\{\phi_\uparrow,\phi_\downarrow\}$ at finite in-plane momentum $\mathbf k$ gives the massless surface Hamiltonian
\begin{equation}
    h_0=\hbar v_\Delta\,(k_x\sigma_y-k_y\sigma_x),\qquad v_\Delta=\Delta/\hbar k_F,
\label{eq:h0plane}
\end{equation}
where the Pauli matrices $\sigma_i$ act on the doublet. Although the doublet is labeled by the spin $s$, the $\sigma_i$ are not the spin operators. Projected onto Eq.~\eqref{eq:phis}, the in-plane components
$s_x$ and $s_y$ vanish, because $(\langle+|-is\langle-|)(|+\rangle+is'|-\rangle)=1+ss'=0$ for $s\neq s'$, while $s_z$ maps onto $\sigma_z$. This is the Ising character of the Majorana Kramers pair~\cite{Chung2009}, and it reduces the exchange coupling to
\begin{equation}
    h_{\rm ex}=\lambda\sum_iS_i^z\,\sigma_z\,\delta(\mathbf r-\mathbf R_i).
\label{eq:hexplane}
\end{equation}
The surface Hamiltonian on the plane is thus $h=h_0+h_{\rm ex}$, valid for energies below the bulk gap $\Delta$. In the next section it is transported to the curved surface of a sphere.

\section{Surface Hamiltonian on the sphere}

In terms of the flat-space gamma matrices $(\gamma^0,\gamma^1,\gamma^2)=(-i\sigma_z,\sigma_x,\sigma_y)$, which satisfy $\{\gamma^i,\gamma^j\}=2\eta^{ij}$ with $\eta_{ij}={\rm diag}(-1,1,1)$, the planar surface Hamiltonian of Eqs.~\eqref{eq:h0plane} and \eqref{eq:hexplane} reads
\begin{equation}
    h=\hbar v_\Delta\,\gamma^0\left(-i\gamma^1\partial_1-i\gamma^2\partial_2\right)
+i\gamma^0\lambda\sum_iS_i^z\,\delta(\mathbf r-\mathbf R_i),
\label{eq:hflat}
\end{equation}
where $i\gamma^0=\sigma_z$ reproduces Eq.~\eqref{eq:hexplane}. The many-body Hamiltonian is $H=\int d^2r\,\Psi^\dagger(\mathbf r)\,h\,\Psi(\mathbf r)$, with $\Psi$ the two-component surface field in the Majorana representation.

On a curved surface the flat metric $\eta_{ij}$ is replaced by $g_{\mu\nu}$, and the gamma matrices and derivatives are transported with the dreibein and the spin connection~\cite{Wald1984,Nakahara1990}. For the space-time $S^2\times\mathbb R$ of a sphere of radius $R$, $g_{\mu\nu}={\rm diag}(-1,R^2,R^2\sin^2\theta)$, the dreibein is $e^i_{~\mu}\equiv\partial x^i/\partial x^\mu={\rm diag}(1,R,R\sin\theta)$, and its inverse is $e_i^{~\mu}={\rm diag}(1,1/R,1/(R\sin\theta))$. The curved-space gamma matrices $\gamma^\mu(x)=\gamma^ie_i^{~\mu}$ are therefore $\gamma^0(x)=-i\sigma_z$, $\gamma^1(x)=\sigma_x/R$, and $\gamma^2(x)=\sigma_y/(R\sin\theta)$. Parallel transport replaces $\partial_\mu$ by the covariant derivative $D_\mu=\partial_\mu+\Gamma_\mu$ with the spin connection $\Gamma_\mu=-\frac{i}{2}\sigma_z\cos\theta\,\delta_{\mu,2}$, whose role is to make the Dirac operator Hermitian with respect to the invariant measure $\sqrt{|g|}\,d^2x=R^2\sin\theta\,d\theta\,d\phi$. The surface Hamiltonian density on the sphere, $h^{\rm cur}=h_0^{\rm cur}+h_{\rm ex}^{\rm cur}$, then reads
\begin{equation}
\begin{aligned}
    h_0^{\rm cur}&=\frac{\Delta}{k_F}\,\gamma^0\left[-i\gamma^1(x)D_1-i\gamma^2(x)D_2\right]\\
    &=\frac{\Delta}{k_FR}\,\gamma^0\left[-i\sigma_x\left(\partial_\theta+\frac{\cot\theta}{2}\right)
    -i\frac{\sigma_y}{\sin\theta}\,\partial_\phi\right],\\
    h_{\rm ex}^{\rm cur}&=i\gamma^0\lambda\sum_iS_i^r\,\delta(\mathbf x-\mathbf R_i)\frac{1}{\sqrt{|g|}},
\end{aligned}
\label{eq:hcur}
\end{equation}
where the $\cot\theta/2$ term comes from the spin connection through $\sigma_y\sigma_z=i\sigma_x$. In the exchange term, $S_i^r$ is the radial component of $\mathbf S_i$, the component normal to the tangent plane at $\mathbf R_i$, which is the one selected by the Ising property, and $g$ is the determinant of $g_{\mu\nu}$, so that
$\delta(\mathbf x-\mathbf R_i)/\sqrt{|g|}$ is the covariant delta function. The many-body Hamiltonian on $S^2\times\mathbb R$ is
\begin{equation}
    H^{\rm cur}=\int_{S^2}d^2x\sqrt{|g|}\,\sum_{\alpha,\beta}\Psi^\dagger_\alpha\,h^{\rm cur}_{\alpha\beta}\,\Psi_\beta,
\label{eq:Hcur}
\end{equation}
and the fields obey the curved-space anticommutation relations
\begin{align}
    \{\Psi_\alpha(\mathbf x,t),\Psi_\beta(\mathbf x',t)\}&=\{\Psi^\dagger_\alpha(\mathbf x,t),\Psi^\dagger_\beta(\mathbf x',t)\}=0,\nonumber\\
    \{\Psi^\dagger_\alpha(\mathbf x,t),\Psi_\beta(\mathbf x',t)\}&=\frac{1}{\sqrt{|g|}}\,\delta_{\alpha\beta}\,\delta^2(\mathbf x-\mathbf x').
\label{eq:anticomm}
\end{align}

\section{Eigenstates and Green function on the sphere}

The imaginary-time Green function of the field in Eq.~\eqref{eq:Hcur}, $G_{\alpha\beta}(\tau,\mathbf x,\mathbf x')=-\langle T_\tau\Psi_\alpha(\tau,\mathbf x)\Psi^\dagger_\beta(0,\mathbf x')\rangle$,
satisfies
\begin{equation}
    \left(-\partial_\tau-h^{\rm cur}\right)G(\tau,\mathbf x,\mathbf x')
=\frac{1}{\sqrt{|g|}}\,\delta(\tau)\,\delta^2(\mathbf x-\mathbf x')\,\mathbf 1,
\label{eq:Geq}
\end{equation}
with $\mathbf 1$ the $2\times2$ identity. The metric factor is removed by the rescaling
\begin{equation}
    \tilde\Psi=|g|^{1/4}\,\Psi,
\label{eq:rescale}
\end{equation}
after which the field obeys the flat-space relation $\{\tilde\Psi^\dagger_\alpha(\mathbf x,\tau),\tilde\Psi_\beta(\mathbf x',\tau)\}=\delta_{\alpha\beta}\,\delta^2(\mathbf x-\mathbf x')$ and the Hamiltonian becomes
\begin{equation}
\begin{aligned}
    H^{\rm cur}&=\int_{S^2}d^2x\sum_{\alpha,\beta}\tilde\Psi^\dagger_\alpha\,\tilde h^{\rm cur}_{\alpha\beta}\,\tilde\Psi_\beta,\\
    \tilde h^{\rm cur}&=h_0^{\rm cur}+i\gamma^0\,\mathcal M(\mathbf x),\qquad
    \mathcal M(\mathbf x)=\lambda\sum_iS_i^r\,\delta(\mathbf x-\mathbf R_i).
\end{aligned}
\label{eq:Htilde}
\end{equation}
In Matsubara frequency the Green function is $G(iw)=(iw-\tilde h^{\rm cur})^{-1}$, and the bare one, which enters the
RKKY calculation, is $G_0(iw)=(iw-h_0^{\rm cur})^{-1}$.

To diagonalize $h_0^{\rm cur}$ we start from the eigenfunctions of the spatial Dirac operator $\mathcal D\equiv-i\gamma^1(x)D_1-i\gamma^2(x)D_2$ in Eq.~\eqref{eq:hcur}. They are labeled by a half-integer angular momentum $l$, its projection $m$ along the polar axis, $|m|\le l$, and $\sigma=\pm1$, with $\mathcal D\,\Upsilon^\sigma_{lm}=\sigma(l+1/2)R^{-1}\,\Upsilon^\sigma_{lm}$~\cite{AbrikosovJr}. With $l^\pm=l\pm1/2$, $m^\pm=m\pm1/2$, and $u=\cos\theta$, they read
\begin{equation}
\begin{aligned}
    \Upsilon^\sigma_{lm}(\theta,\phi)&=\mathcal N_{lm}\,\frac{e^{im\phi}}{\sqrt{2\pi}\,R}
    \begin{pmatrix} f^{-}_{lm}(u)\\[2pt] i\sigma\,{\rm sgn}(m)\,f^{+}_{lm}(u)\end{pmatrix},\\[4pt]
    f^{\mp}_{lm}(u)&=\sqrt{\rho^{(|m^\mp|,|m^\pm|)}(u)}\;P^{(|m^\mp|,|m^\pm|)}_{l-|m|}(u),\\[4pt]
    \mathcal N_{lm}&=i^{l^+}(-1)^{\frac12(m^-+|m^-|)}
    \frac{\sqrt{(l+m)!\,(l-m)!}}{2^{|m|+\frac12}\,\Gamma(l^+)},
\end{aligned}
\label{eq:Upsilon}
\end{equation}
where $P^{(\alpha,\beta)}_n$ are Jacobi polynomials, $\rho^{(\alpha,\beta)}(u)=(1-u)^\alpha(1+u)^\beta$ their weight function, and the components refer to the rescaled field $\tilde\Psi$. Since the two components differ only by the sign of $\sigma$ in the lower entry,
\begin{equation}
    \sigma_z\,\Upsilon^\sigma_{lm}=\Upsilon^{-\sigma}_{lm}.
\label{eq:sigmaz}
\end{equation}
The Hamiltonian $h_0^{\rm cur}=(\Delta/k_F)\gamma^0\mathcal D$ is therefore off-diagonal in $\sigma$. Acting with $\gamma^0=-i\sigma_z$,
\begin{equation}
    h_0^{\rm cur}\,\Upsilon^\sigma_{lm}=-i\sigma\,\mathcal E_l\,\Upsilon^{-\sigma}_{lm},\qquad
    \mathcal E_l=\frac{\Delta}{k_FR}\left(l+\frac12\right),
\label{eq:h0Upsilon}
\end{equation}
so that its eigenstates are the combinations
\begin{equation}
    |\Psi^\pm_{lm}\rangle=\frac{1}{\sqrt2}\left(|\Upsilon^+_{lm}\rangle\mp i|\Upsilon^-_{lm}\rangle\right),\qquad
    h_0^{\rm cur}|\Psi^\pm_{lm}\rangle=\pm\mathcal E_l|\Psi^\pm_{lm}\rangle,
\label{eq:Psi}
\end{equation}
and the bare Green function is diagonal in this basis,
\begin{equation}
    \langle\Psi^\pm_{lm}|G_0(iw)|\Psi^\pm_{lm}\rangle=\frac{1}{iw\mp\mathcal E_l}.
\label{eq:G0}
\end{equation}

\section{RKKY interaction on the sphere}

Consider two impurities at $\mathbf R_1$ and $\mathbf R_2$. Integrating
out the Majorana modes to second order in $\lambda$ gives
\begin{equation}
    H^{\rm RKKY}_{1,2}=\lambda^2\chi_{1,2}\,S^r_1S^r_2,
\label{eq:HRKKY12}
\end{equation}
where $\chi_{1,2}$ is the static limit of $\chi_{1,2}(\tau)=\langle T_\tau s^r(\mathbf R_1,\tau)\,s^r(\mathbf R_2,0)\rangle$, with $s^r(\mathbf R)=\tilde\Psi^\dagger(\mathbf R)\sigma_z\tilde\Psi(\mathbf R)$ the radial spin density of the surface modes. At one loop and at bosonic Matsubara frequency $\nu_n$,
\begin{equation}
\begin{aligned}
    \chi_{1,2}(i\nu_n)=-T\sum_{iw_n}\Tr\big[&s^r_1\,G_0(\mathbf R_1,\mathbf R_2,i\nu_n+iw_n)\\
    &\times s^r_2\,G_0(\mathbf R_2,\mathbf R_1,iw_n)\big],
\end{aligned}
\label{eq:chi_loop}
\end{equation}
with $w_n=(2n+1)\pi T$, $k_B\equiv1$, and the trace over the eigenbasis $|\Psi^\pm_{lm}\rangle$ of Eq.~\eqref{eq:Psi}. The static susceptibility, $\nu_n=0$, reads
\begin{equation}
\begin{aligned}
    \chi_{1,2}=-T\sum_{iw_n}\sum_{a,lm}\sum_{b,l'm'}
    &\langle\Psi^a_{lm}|s^r_1|\Psi^b_{l'm'}\rangle
    \langle\Psi^b_{l'm'}|G_0|\Psi^b_{l'm'}\rangle\\
    \times&\langle\Psi^b_{l'm'}|s^r_2|\Psi^a_{lm}\rangle
    \langle\Psi^a_{lm}|G_0|\Psi^a_{lm}\rangle.
\end{aligned}
\label{eq:chi_static}
\end{equation}
Since $\langle\tilde\Psi_\alpha(\mathbf R_i)|s^r_i|\tilde\Psi_\beta(\mathbf R_i)\rangle=(\sigma_z)_{\alpha\beta}$, Eqs.~\eqref{eq:Upsilon}, \eqref{eq:sigmaz}, and \eqref{eq:Psi} give the matrix elements of $s^r_i$ as a $2\times2$ matrix in the band index $a,b=\pm$,
\begin{equation}
\begin{aligned}
    \langle\Psi^a_{lm}|s^r_i|\Psi^b_{l'm'}\rangle
    =&\,\frac{e^{-i(m-m')\phi_i}}{2\pi R^2}\,\mathcal N^*_{lm}\mathcal N_{l'm'}\\
    &\times\left[B_{lm;l'm'}(u_i)\,\mathbf 1-C_{lm;l'm'}(u_i)\,\sigma_y\right]_{ab},
\end{aligned}
\label{eq:sr_matrix}
\end{equation}
with $u_i=\cos\theta_i$ and
\begin{equation}
\begin{aligned}
    B_{lm;l'm'}(u)&=f^-_{lm}(u)f^-_{l'm'}(u)-{\rm sgn}(mm')\,f^+_{lm}(u)f^+_{l'm'}(u),\\
    C_{lm;l'm'}(u)&=f^-_{lm}(u)f^-_{l'm'}(u)+{\rm sgn}(mm')\,f^+_{lm}(u)f^+_{l'm'}(u).
\end{aligned}
\label{eq:BC}
\end{equation}
The diagonal part $B$ connects states of the same band and the off-diagonal part $C$ connects the positive and negative energy bands. Inserting Eqs.~\eqref{eq:G0} and \eqref{eq:sr_matrix} into Eq.~\eqref{eq:chi_static},
\begin{equation}
\begin{aligned}
    \chi_{1,2}=-T\sum_{iw_n}\sum_{lm,l'm'}&\frac{|\mathcal N_{lm}\mathcal N_{l'm'}|^2}{4\pi^2R^4}\,
    e^{-i(m-m')(\phi_1-\phi_2)}\\
    &\times\left[F_1+F_2-F_3-F_4\right],
\end{aligned}
\label{eq:chi_F}
\end{equation}
where, with $B_{1,2}\equiv B_{lm;l'm'}(u_{1,2})$, $C_{1,2}\equiv C_{lm;l'm'}(u_{1,2})$,
\begin{equation}
\begin{aligned}
    F_1&=\frac{B_1B_2}{(iw_n+\mathcal E_l)(iw_n+\mathcal E_{l'})},\\
    F_2&=\frac{B_1B_2}{(iw_n-\mathcal E_l)(iw_n-\mathcal E_{l'})},\\
    F_3&=\frac{C_1C_2}{(iw_n+\mathcal E_l)(iw_n-\mathcal E_{l'})},\\
    F_4&=\frac{C_1C_2}{(iw_n-\mathcal E_l)(iw_n+\mathcal E_{l'})}.
\end{aligned}
\label{eq:F1234}
\end{equation}
The Matsubara sum is evaluated by contour integration,
\begin{equation}
\begin{aligned}
    -T\sum_{iw_n}&\left[F_1+F_2-F_3-F_4\right]\\
    =&-\frac{B_1B_2}{\mathcal E_l-\mathcal E_{l'}}
    \Big(f[\mathcal E_l]-f[\mathcal E_{l'}]+f[-\mathcal E_{l'}]-f[-\mathcal E_l]\Big)\\
    &+\frac{C_1C_2}{\mathcal E_l+\mathcal E_{l'}}
    \Big(f[\mathcal E_l]-f[-\mathcal E_{l'}]+f[\mathcal E_{l'}]-f[-\mathcal E_l]\Big),
\end{aligned}
\label{eq:matsubara}
\end{equation}
with $f$ the Fermi function. The spectrum of Eq.~\eqref{eq:Psi} has a gap $2\Delta/(k_FR)$ from the spin connection. In the low-temperature limit $T\ll\Delta/(k_FR)$ the intraband terms in the first line vanish, and the interband terms give
\begin{equation}
\begin{aligned}
    \chi_{1,2}=-\frac{1}{2\pi^2R^4}\sum_{lm,l'm'}&\,e^{-i(m-m')(\phi_1-\phi_2)}\,|\mathcal N_{lm}\mathcal N_{l'm'}|^2\\
    &\times\frac{C_{lm;l'm'}(u_1)\,C_{lm;l'm'}(u_2)}{\mathcal E_l+\mathcal E_{l'}},
\end{aligned}
\label{eq:chi_lowT}
\end{equation}
which is Eq.~\eqref{eq:susceptiblity} of the main text. The sum runs over $|m|\le l\le l_{\rm max}+1/2$, where the integer $l_{\rm max}\simeq k_FR$ marks the angular momentum at which $\mathcal E_l$ reaches the bulk gap $\Delta$. By rotational symmetry we may set $\phi_1=\phi_2$, place one impurity at the north pole and the other at polar angle $\theta$, and evaluate Eq.~\eqref{eq:chi_lowT} as a function of $k_Fd=l_{\rm max}\theta$. The result is shown in Fig.~\ref{fig:RKKY_oscillation} of the main text and is negative at all separations, so the RKKY coupling $\lambda^2\chi_{1,2}$ is ferromagnetic in the radial components.

\subsection{Asymptotic form}

For the asymptotic form we place both impurities on the equator, $\theta_1=\theta_2=\pi/2$, at azimuthal separation $\phi$. By rotational symmetry this is equivalent to the configuration of Fig.~\ref{fig:RKKY_oscillation} of the main text, one impurity at the north pole and the other at polar angle $\theta=\phi$, since only the arc distance between the two impurities enters. With $u_1=u_2=0$ and $\mathcal E_l+\mathcal E_{l'}=(l+l'+1)\Delta/(k_FR)$,
\begin{equation}
\begin{aligned}
    \chi(\phi)=-\frac{k_F}{2\pi^2R^3\Delta}\sum_{lm,l'm'}&\,e^{-i(m-m')\phi}\,|\mathcal N_{lm}\mathcal N_{l'm'}|^2\\
    &\times\frac{C_{lm;l'm'}(0)^2}{l+l'+1}.
\end{aligned}
\label{eq:chi_equator}
\end{equation}
Writing $l=n+1/2$ and $m=k+1/2$ with integers $n$ and $k=0,\dots,n$, and using the properties of the Jacobi polynomials, the sums over $m,m'$ can be organized into the two functions
\begin{equation}
\begin{aligned}
    F_n(\phi)&=\sum_{k=0}^n c_{n,k}\cos(k_+\phi),\\
    G_n(\phi)&=\sum_{k=0}^n(-1)^{n-k}\,c_{n,k}\sin(k_+\phi),\\
    c_{n,k}&=\frac{(n+k+1)!\,(n-k)!}{(n!)^2\,2^{2k+1}}\left(P^{(k,k+1)}_{n-k}(0)\right)^2,
\end{aligned}
\label{eq:FG}
\end{equation}
with $k_+=k+1/2$, so that, with the integer cutoff $l_{\rm max}$ of Eq.~\eqref{eq:chi_lowT}, where $n\le l_{\rm max}$ corresponds to
$l\le l_{\rm max}+1/2$,
\begin{equation}
    \chi(\phi)=-\frac{k_F}{\pi^2R^3\Delta}\sum_{n,n'=0}^{l_{\rm max}}
    \frac{F_n(\phi)F_{n'}(\phi)+G_n(\phi)G_{n'}(\phi)}{n+n'+2}.
\label{eq:chi_FG}
\end{equation}
The remaining steps are a rough estimate whose purpose is to arrive at a closed form, which is then checked against the exact sums. Using $P^{(k,k+1)}_{n-k}(0)=P^{(k,k)}_{n-k}(0)+\frac{n-k+1}{n+1}P^{(k,k)}_{n-k+1}(0)$
and the fact that $P^{(k,k)}_n(0)$ is nonzero only for even $n$, where it is expressible through factorials, one finds for even $n$
\begin{equation}
    F_n(\phi)=\frac1\pi\sum_{k=0}^n\frac{\eta_{n,k}\,(n+k+1)!}{(n-k)!\,2^{2k+1}}\cos(k_+\phi),
\label{eq:Fn_eta}
\end{equation}
with $\eta_{n,k}=[\Gamma((n-k+1)/2)/\Gamma((n+k)/2+1)]^2$ for even $k$ and $\eta_{n,k}=[\Gamma((n-k)/2+1)/\Gamma((n+k)/2+1)]^2$ for odd $k$, that is,
\begin{equation}
\begin{aligned}
    F_n(\phi)=&\sum_{k\ {\rm even}}^n a_{n,k}\cos(k_+\phi)
    +\sum_{k\ {\rm odd}}^n b_{n,k}\cos(k_+\phi),\\
    a_{n,k}=&\,\frac{(n+k+1)!\,(n-k)!}{2^{2n+1}\left[((n+k)/2)!\,((n-k)/2)!\right]^2},\\
    b_{n,k}=&\,\frac{(n+k+1)!\,(n-k+1)!\,(n-k+1)}{2^{2n+3}\left[((n+k+1)/2)!\,((n-k+1)/2)!\right]^2}.
\end{aligned}
\label{eq:Fn_evenodd}
\end{equation}
Extending $a_{n,k}$ to all $k$, with an overall factor $1/2$ to compensate the resulting double counting, gives the estimate
\begin{equation}
\begin{aligned}
    F_n(\phi)&\simeq\frac12\sum_{k=0}^n a_{n,k}\cos(k_+\phi),\\
    G_n(\phi)&\simeq\frac12\sum_{k=0}^n a_{n,k}\sin(k_+\phi),
\end{aligned}
\label{eq:FG_approx}
\end{equation}
where in $G_n$ the oscillating factor $(-1)^{n-k}$ has been dropped as for even $n$ and $k$. Stirling's formula, $(n+k)!/[((n+k)/2)!]^2\simeq2^{n+k+1}/\sqrt{2\pi(n+k)}$, then gives
\begin{equation}
\begin{aligned}
    F_n(\phi)&\simeq\frac{1}{2\pi}\sum_{k=0}^{n-1}\frac{n+k+1}{\sqrt{n^2-k^2}}\cos(k_+\phi),\\
    G_n(\phi)&\simeq\frac{1}{2\pi}\sum_{k=0}^{n-1}\frac{n+k+1}{\sqrt{n^2-k^2}}\sin(k_+\phi),
\end{aligned}
\label{eq:FG_stirling}
\end{equation}
which extend to odd $n$. In the continuum limit, matching the amplitude at $\phi=0$,
\begin{equation}
    F_x(\phi)=\frac{1}{2+\pi}\int_0^x dy\sqrt{\frac{x+y}{x-y}}\cos(y\phi),\qquad x\simeq n.
\label{eq:F_cont}
\end{equation}
These estimates identify the Bessel functions as the natural closed form. For large $n$ and $\phi\ll\pi$,
\begin{equation}
\begin{aligned}
    F_n(\phi)&\simeq\frac{n+1}{2}J_0\!\left((n+\tfrac12)\phi\right),\\
    G_n(\phi)&\simeq\frac{n+1}{2}J_1\!\left((n+\tfrac12)\phi\right),
\end{aligned}
\label{eq:FG_bessel}
\end{equation}
which we verified against the exact sums of Eq.~\eqref{eq:FG}. The identity $F_n(0)=(n+1)/2$ holds exactly, and for $\phi\le0.2$ the deviations are below a few percent for $n\ge10$ and decrease with $n$. Inserting Eq.~\eqref{eq:FG_bessel} into Eq.~\eqref{eq:chi_FG} and returning to $l=n+1/2$ gives Eq.~\eqref{eq:RKKY asymptotic} of the main text,
\begin{equation}
\begin{aligned}
    \chi(\phi)&=-\frac{k_F}{4\pi^2R^3\Delta}\sum_{l,l'=1/2}^{l_{\rm max}+1/2}
    \frac{(l+1/2)(l'+1/2)}{l+l'+1}\,\mathcal F_{l,l'}(\phi),\\
    \mathcal F_{l,l'}(\phi)&=J_0(l\phi)J_0(l'\phi)+J_1(l\phi)J_1(l'\phi),
\end{aligned}
\label{eq:chi_asym}
\end{equation}
from which the flat-surface result is recovered as $R\to\infty$. Its agreement with the exact Eq.~\eqref{eq:chi_lowT} is shown in the inset of Fig.~\ref{fig:RKKY_oscillation} of the main text.

\section{Transfer matrix on a great circle}

For $N$ impurities equally spaced on a great circle through the poles, $\theta_i=2\pi i/N$, the Hamiltonian of Eq.~\eqref{Eq:Hij} of the main text is a sum of nearest-neighbor bond terms, which we symmetrize as 
\begin{equation}
\begin{aligned}
    H(\Theta_i,\Theta_{i+1})=&-J\cos\Theta_i\cos\Theta_{i+1}\\
    &-\frac{B}{2}\big[\cos(\Theta_i+\theta_i)+\cos(\Theta_{i+1}+\theta_{i+1})\big],
\end{aligned}
\label{eq:bond}
\end{equation}
with $J>0$, so that $H=\sum_{i=1}^NH(\Theta_i,\Theta_{i+1})$ and $\Theta_{N+1}\equiv\Theta_1$. The moments are kept in the meridian plane, as in the main text. With the transfer matrix $T(\Theta_i,\Theta_{i+1})=e^{-H(\Theta_i,\Theta_{i+1})/T}$, the partition function is
\begin{equation}
    Z=\int\prod_{i=1}^N\frac{d\Theta_i}{2\pi}\,
    T(\Theta_1,\Theta_2)\,T(\Theta_2,\Theta_3)\cdots T(\Theta_N,\Theta_1).
\label{eq:Z_int}
\end{equation}
Discretizing $\Theta_i=2\pi n_i/M$ with $n_i=1,\dots,M$ turns each factor into an $M\times M$ matrix $T_{i,i+1}$ with elements $T(2\pi n/M,2\pi n'/M)$ and the integral into a cyclic matrix product,
\begin{equation}
    Z=M^{-N}\,\tilde Z,\qquad
    \tilde Z=\Tr\left[T_{1,2}\,T_{2,3}\cdots T_{N,1}\right].
\label{eq:Z_tm}
\end{equation}
Because the bond terms depend on the positions $\theta_i$, the $N$ matrices are all different and the product is evaluated numerically. We use $M=100$, for which the results are converged. The axial magnetization $m_{\rm ax}=-\frac1N\partial F/\partial B$, with
$F=-T\ln Z$, follows by differentiating the product. Each bond contributes $\frac{1}{2T}(m_i+m_{i+1})$ under the derivative, where $m_i={\rm diag}[\cos(2\pi n/M+\theta_i)]$, so that each site appears twice and
\begin{equation}
\begin{aligned}
    m_{\rm ax}=\frac{1}{N\tilde Z}\sum_{i=1}^N
    \Tr\big[&T_{1,2}\cdots T_{i-1,i}\,m_i\\
    &\times T_{i,i+1}\cdots T_{N,1}\big].
\end{aligned}
\label{eq:max_tm}
\end{equation}
Figure~\ref{fig:magnetization} of the main text shows $m_{\rm ax}$ versus $B$ at $T/J=0.1$ for $N=40$ to $52$ and the resulting critical field.

\section{Icosahedral triangulation of the sphere}

The impurity positions on the sphere are the vertices of a class-I icosahedral geodesic triangulation. Each of the 20 faces of a regular icosahedron is divided into $\nu^2$ equal triangles by splitting its edges into $\nu$ segments, and the resulting vertices are projected radially onto the sphere. The icosahedron is oriented with two of its vertices at the poles, so that the lattice is inversion symmetric. Its remaining ten vertices lie in two rings of five at $z=\pm1/\sqrt5$, with the ten faces between the rings forming a band around the equator. For a subdivision into $\nu$ segments per edge the triangulation has
\begin{equation}
    N=10\nu^2+2,\qquad N_e=30\nu^2,\qquad N_t=20\nu^2
\label{eq:counts}
\end{equation}
vertices, edges, and triangles, consistent with Euler's relation $N-N_e+N_t=2$ and with $N_e=3(N-2)$, $N_t=2(N-2)$ for any triangulation of the sphere. The twelve icosahedron vertices have five neighbors and all others six.

The nearest-neighbor spacing follows from the area per triangle. For a sphere of radius $R$, $A_t=4\pi R^2/N_t$, and approximating each triangle by a planar equilateral one of side $d$, $A_t=\sqrt3d^2/4$, gives the arc angle
\begin{equation}
    \theta_d=\frac dR=\sqrt{\frac{16\pi}{\sqrt3N_t}}\simeq\frac{3.8}{\sqrt N},
\label{eq:thetad}
\end{equation}
which is used in the main text. The exact spherical value is obtained from the spherical excess, $A_t=R^2(3\alpha_t-\pi)$ with $\alpha_t$ the angle of the equilateral spherical triangle, so that $\alpha_t=\pi(4/N_t+1)/3$ and $\cos\theta_d=\cos\alpha_t(1+\cos\alpha_t)/\sin^2\alpha_t$ by the spherical law of cosines. The two agree to better than one percent for $N\gtrsim100$.

The number of bonds cut by the equator enters the domain-wall energy of Eq.~\eqref{eq:E_sphere} of the main text. For odd $\nu$ no vertex lies on the equator, and the equator crosses exactly $10\nu=\sqrt{10(N-2)}\simeq\sqrt{10N}$ edges of the equatorial band. For even $\nu$, a ring of $5\nu$ vertices lies exactly on the equator, where $\cos\theta_i=0$ and the sign of $S^r_i$ in the domain-wall configuration is undefined. The Monte Carlo simulations therefore use odd $\nu$, $\nu=5,7,9,11$, corresponding to $N=252,492,812,1212$.

\section{Monte Carlo simulations}

We simulate the Hamiltonian of Eq.~\eqref{Eq:Hij} of the main text,
\begin{equation}
    H=-J\sum_{\langle ij\rangle}S^r_iS^r_j-B\sum_iS^z_i,
\label{eq:H_MC}
\end{equation}
on the icosahedral triangulation of the previous section, with the moments $\mathbf S_i$ as unit vectors ($S$ absorbed into $J$ and $B$), $S^r_i=\mathbf S_i\cdot\hat r_i$, $S^z_i=\mathbf S_i\cdot\hat z$, and the sum over the $N_e$ edges. In the planar version the moments are confined to the meridian plane, $\mathbf S_i=\cos\Theta_i\,\hat r_i+\sin\Theta_i\,\hat\theta_i$, which is exactly Eq.~\eqref{Eq:Hij} of the main text. We also simulated moments free to point anywhere on the unit sphere, which includes fluctuations out of the meridian plane.

\subsection{Sampling}

A Monte Carlo sweep consists of one Metropolis update per site followed by three global moves, which reverse all moments in the southern hemisphere, in the northern hemisphere, and on the whole sphere, respectively, each attempted once per sweep. The first two are inversion images of each other. Inversion exchanges the hemispheres and, since the moments are axial vectors while the outward normal is odd, maps the radial component as $S^r(\mathbf r)\to-S^r(-\mathbf r)$, so it takes the outward hedgehog to the inward one and leaves the domain wall invariant. The single-site proposal is $\Theta_i\to\Theta_i+\sigma\xi$ in the planar case and $\mathbf S_i\to(\mathbf S_i+\sigma\boldsymbol\xi)/|\mathbf S_i+\sigma\boldsymbol\xi|$
otherwise, with $\xi$ a unit Gaussian variable and step width $\sigma=0.15+0.8\sqrt{T/J}$. Both proposals are symmetric, with the probability of proposing a move equal to that of proposing its reverse, so the Metropolis acceptance below needs no correction for the proposal. The energy change involves only the neighbor field $h_i=\sum_{j\in{\rm nbr}(i)}S^r_j$, 
\begin{equation}
    \Delta E_i=-J\,(S^{r\prime}_i-S^r_i)\,h_i-B\,(S^{z\prime}_i-S^z_i),
\label{eq:dE_site}
\end{equation}
and the move is accepted with probability $\min(1,e^{-\Delta E_i/T})$. Since $\Delta E_i$ depends on the neighbors of $i$ only, sites that share no bond can be updated simultaneously without changing the stationary distribution. We therefore partition the sites into classes containing no bonded pair, assigning to each site in turn the lowest label not yet used by any of its neighbors, which yields five classes for the icosahedral triangulation, and update one class at a time with the neighbor fields recomputed in between.

The global moves reverse all moments in a set of sites $\mathcal A$, $\mathbf S_i\to-\mathbf S_i$ for $i\in\mathcal A$. Only bonds with exactly one end in $\mathcal A$ change sign, so
\begin{equation}
    \Delta E_{\mathcal A}=2J\!\!\sum_{\langle ij\rangle\in\partial\mathcal A}\!\!S^r_iS^r_j
    +2B\sum_{i\in\mathcal A}S^z_i,
\label{eq:dE_flip}
\end{equation}
where $\partial\mathcal A$ denotes the bonds crossing the boundary of $\mathcal A$. The move is its own inverse, so the Metropolis acceptance $\min(1,e^{-\Delta E_{\mathcal A}/T})$ satisfies detailed balance. These
moves connect the hedgehog and domain-wall sectors directly. Without them, single-site updates would have to cross a barrier of order $2\sqrt{10N}J$, which exceeds $100J$ for $N\ge252$ and is three orders of magnitude larger than $T=0.1J$, so the simulation would remain in whichever sector it started in.

\begin{figure}[t]
    \centering
    \includegraphics[width=\columnwidth]{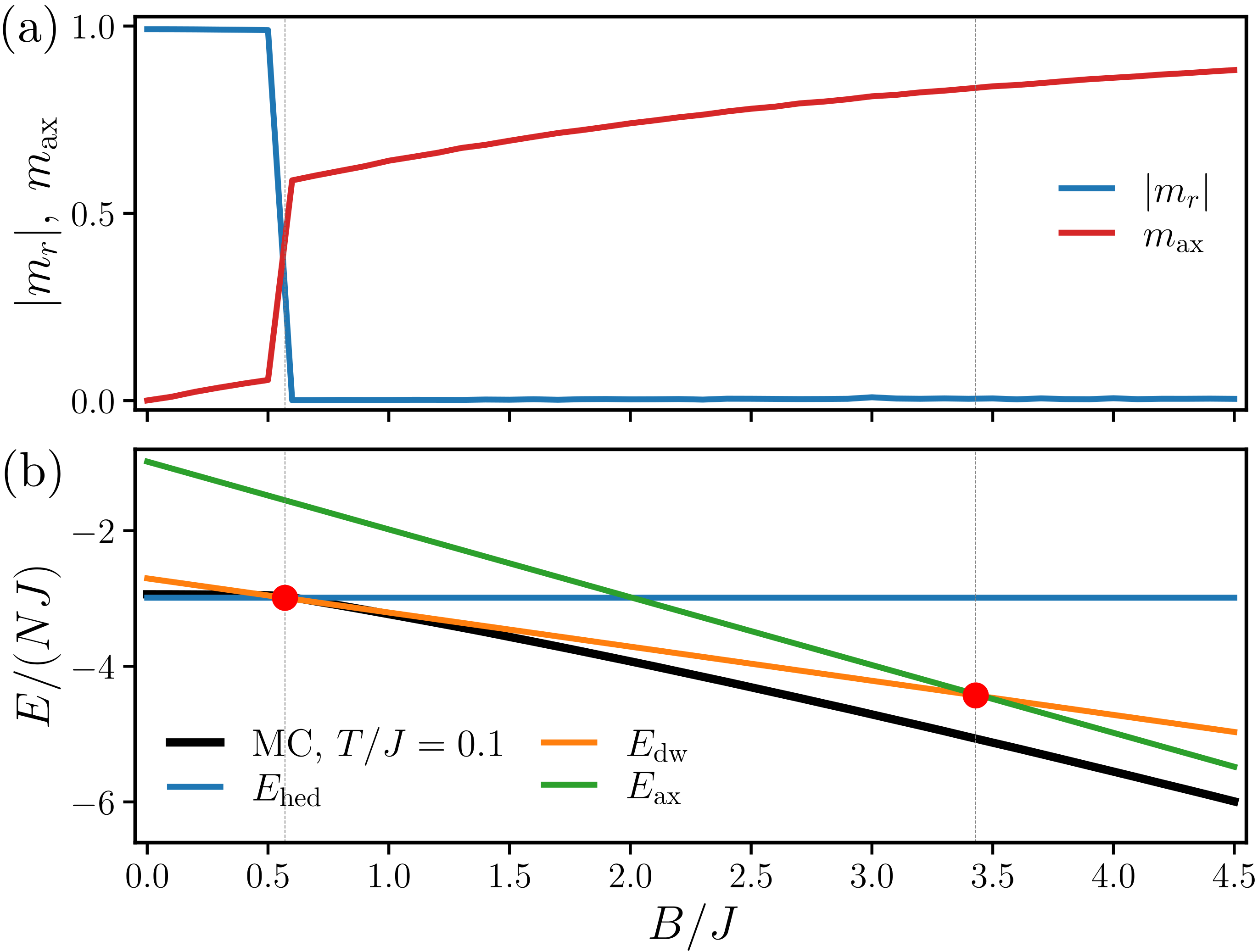}
    \caption{Monte Carlo results for $N=492$ at $T/J=0.1$ over the extended field range. (a) Radial magnetization $|m_r|$ and axial magnetization $m_{\rm ax}$ versus $B/J$. (b) Energy per impurity from the simulation, compared with $E_{\rm hed}$, $E_{\rm dw}$, and $E_{\rm ax}$ of Eq.~\eqref{eq:E_sphere} of the main text. The dots mark the crossings $E_{\rm hed}=E_{\rm dw}$ at $B_c=4J\sqrt{10/N}$ and $E_{\rm dw}=E_{\rm ax}$ at $4J(1-\sqrt{10/N})$, and the thin vertical lines carry these fields to (a). The magnetizations jump at $B_c$ and vary smoothly through the second crossing.}
    \label{fig:MC_largeB}
\end{figure}

\subsection{Protocol and critical field}

At each field $B$ we run three independent simulations, one annealed from a random configuration, from $T=2J$ to $T=0.1J$ in 30 geometric steps of $40+N/20$ sweeps each, and two started from the hedgehog and the domain-wall configurations and cooled from $T=0.4J$ to $T=0.1J$ in 10 steps of 30 sweeps. Each is then measured over 100 sweeps at $T/J=0.1$, and the run with the lowest energy is retained. The ordered starts guard against metastable multidomain states, which annealing occasionally produces at the larger sizes. At $T/J=0.1$ the energies of the two ordered sectors differ by an amount of order $N(B-B_c)$, while their entropies differ by an amount of order one, so selecting by energy and by free energy agree to better than one percent in $B_c$. 

The order parameters are the radial magnetization $m_r=\frac1N\sum_i\langle S^r_i\rangle$ and the axial magnetization $m_{\rm ax}=\frac1N\sum_i\langle S^z_i\rangle$, averaged over the measurement sweeps. With the unit-vector normalization used here these coincide with $m_r=\frac{1}{NS}\sum_i\langle S^r_i\rangle$ of the main text and with $m_{\rm ax}=-\frac1N\partial F/\partial B$ of the transfer-matrix calculation. We record $|m_r|$ because the two orientations of the hedgehog are degenerate at $B=0$ and the whole-sphere reversal moves between them. The hedgehog has $|m_r|=1$ and $m_{\rm ax}\simeq0$, the domain wall $|m_r|=0$ and $m_{\rm ax}\simeq\langle|\cos\theta|\rangle=1/2$. The critical field at $T/J=0.1$ is defined by $|m_r|=1/2$, interpolated linearly between field points spaced by $0.05J$. The transition is sharp, so the uncertainty is below half a grid step.

As an independent estimate we also determine $B_c$ at zero temperature. Starting from the hedgehog and from the domain wall, each configuration is cooled from $T=0.3J$ to $T=0.02J$ with the global moves disabled, so that the moments relax without leaving their sector. The field at which the two relaxed energies cross is $B_c(T\to0)$. The relaxation consists of a small tilt of each moment toward the field, which Eq.~\eqref{eq:E_sphere} of the main text ignores. For a moment at polar angle $\theta_i$ the tilt is of order $B\sin\theta_i/(zJ)$, with $z$ the coordination number, and lowers the energy by about $B^2\sin^2\theta_i/(2zJ)$. This gain depends on $\sin^2\theta_i$ only, not on the sign of $S^r_i$, so the hedgehog and the domain wall gain the same amount and the crossing field is unchanged at this order. Figure~\ref{fig:Configuration}(c) of the main text confirms this, the two estimates agreeing within a few percent.

Simulations with the moments free to leave the meridian plane give the same $B_c$ within one grid step at $N=492$. The out-of-plane component
couples to neither term of Eq.~\eqref{eq:H_MC}, so it affects only the entropy, which is negligible at $T/J=0.1$.

\subsection{Behavior above the critical field}

Figure~\ref{fig:MC_largeB} extends the simulations at $N=492$ to $B=4.5J$, beyond the crossing of $E_{\rm dw}$ and $E_{\rm ax}$ of Eq.~\eqref{eq:E_sphere} of the main text, which lies at $B\simeq4J(1-\sqrt{10/N})=3.4J$ for this size. Above $B_c$, $m_{\rm ax}$ increases smoothly and $|m_r|$ stays at zero, with no jump at any field. The simulated energy per impurity lies below all three configurations of Eq.~\eqref{eq:E_sphere} of the main text for $B\gtrsim0.9J$, because the moments cant continuously toward the field, and it passes smoothly through the field where $E_{\rm dw}$ and $E_{\rm ax}$ cross. The sharp wall therefore evolves without a transition into the axial texture $S^r\simeq\cos\theta$, which also reverses sign at the equator, as stated in the main text.

\end{document}